# MHD activity induced coherent mode excitation in the edge plasma region of ADITYA-U Tokamak

Kaushlender Singh[1,2], Suman Dolui[1,2], Bharat Hegde[1,2], Lavkesh Lachhvani[1,2], Sharvil Patel[1,4], Injamul Hoque[1,2], Ashok K. Kumawat[1,2], Ankit Kumar[1,2], Tanmay Macwan[3], Harshita Raj[1,2], Soumitra Banerjee[1,2], Komal Yadav[1,2], Abha Kanik[1,6], Pramila Gautam[1], Rohit Kumar[1], Suman Aich[1], Laxmikanta Pradhan[1], Ankit Patel[1], Kalpesh Galodiya[1], Daniel Raju[1,2], S.K. Jha[1], K. A. Jadeja[1,7], K. M. Patel[1], S. N. Pandya[1], M.B. Chaudhary[1], R.L. Tanna[1,5], P. K. Chattopadhyay[1,2], R. Pal[8], Y. C. Saxena[1,2], Abhijit Sen[1,2] and Joydeep Ghosh[1,2]

[1]Institute for Plasma Research, Bhat, Gandhinagar, Gujarat, 382428, India
[2]Homi Bhabha National Institute, Training School Complex, Anushakti Nagar, Mumbai, 400094, India
[3]University of California Los Angeles, Department of Physics and Astronomy Los Angeles, CA, 90095, USA
[4]Pandit Deendayal Petroleum University, Raisan, Gandhinagar, Gujarat, 382007, India
[5]Institute of Science, Nirma University, Ahmedabad, Gujarat, 382481, India
[6]University of Petroleum and Energy Studies, Bidholi, Uttarakhand, 248007, India
[7]Department of Nano Science and Advanced Materials, Saurashtra University, Rajkot, Gujarat, 360005, India
[8]Saha Institute of Nuclear Physics, Kolkata, 700064, India

Email: Kaushlender.singh@ipr.res.in

**Abstract:** In this paper, we report the excitation of coherent density and potential fluctuations induced by magnetohydrodynamic (MHD) activity in the edge plasma region of ADITYA-U Tokamak. When the amplitude of the MHD mode, mainly the m/n = 2/1, increases beyond a threshold value, $|\tilde{B}_\theta|/B_\theta \sim 0.3 - 0.4$ %, coherent oscillations in the density and potential fluctuations are observed having the same frequency as that of the MHD mode. The mode numbers of these MHD induced density and potential fluctuations are obtained by Langmuir probes placed at different radial, poloidal, and toroidal locations in the edge plasma region. Detailed analyses of these Langmuir probe measurements reveal that the coherent mode in edge potential fluctuation has a mode structure of m/n = 2/1 whereas the edge density fluctuation has an m/n = 1/1 structure. It is further observed that beyond the threshold, the coupled power fraction scales almost linearly with the magnitude of $\tilde{B}_\theta/B_\theta$ fluctuations. Furthermore, the rise rates of the coupled power fraction for coherent modes in density and potential fluctuations are also found to be dependent on the growth rate of magnetic fluctuations. The disparate mode structures of the excited modes in density and plasma potential fluctuations suggest that the underlying mechanism for their existence is most likely due to the excitation of the global high-frequency branch of zonal flows occurring through the coupling of even harmonics of potential to the odd harmonics of pressure due to 1/R dependence of the toroidal magnetic field.

## 1 Introduction:

To minimize the cross-field transport in magnetically confined plasmas, it is crucial to understand the underlying dynamics that control various transport processes [1][2][3]. Prediction and control of transport are essential for optimising and enhancing the performance of tokamak devices, including the International Thermonuclear Experimental Reactor (ITER) [4]. The free energy associated with the gradients in temperature, density, and potential ($\nabla T_e$, $\nabla n_e$, and $\nabla V_p$) induce electrostatic turbulence, which is known to be the primary contributor to anomalous transport in a tokamak plasma [1]. The excitation of self-organized coherent structures such as Geodesic Acoustic Modes (GAMs) [5] and low-frequency zonal flows (LFZFs) [6][7], are commonly observed to saturate the turbulence and are known to cause reduction in the related cross-field transport in tokamak plasmas [5][7]. The drift turbulence via turbulent Reynolds stress (RS) or transport modulations and energetic particles (EPs) are known to be driving these coherent structures[5][8][9]. As it is quite well known that the edge plasma region has a substantial influence on the behaviour of the entire plasma column in tokamaks, these modifications in the edge region are also responsible for improvement in global plasma confinement [10].

The poloidal magnetic field oscillations ($\tilde{B}_\theta$) can also modify the edge turbulence to excite coherent oscillations such as zonal flows and other GAM-like oscillations [11][12][13][14]. These $\tilde{B}_\theta$ fluctuations can be generated either due to the growth of internal MHD modes or by externally applied resonant magnetic perturbations (RMP). Since, the interaction of edge plasma turbulence with magnetic oscillations is crucial for understanding the dynamics of turbulence and transport, several experimental and theoretical works have been conducted in the past [15][16]. In the TCABR tokamak, the edge turbulence modulation in the presence of high MHD activity at a frequency around 10 kHz has been experimentally explored. Wavelet cross-spectral analyses have revealed phase locking and frequency synchronization between m/n = 3/1 magnetic island rotation-induced magnetic fluctuations ($\tilde{B}_\theta$) and potential fluctuations [11]. Coherent oscillations in core plasma potential fluctuation induced by high amplitude m = 2 MHD mode has also been reported in T-10 tokamak [17]. Similarly, a dominant peak around 14.5 kHz in the power spectra of core density fluctuations has been observed to be correlated to the rotation frequency of an m/n =2/1 magnetic island in the DIII-D tokamak [18]. In the HL-2A tokamak the m/n = 2/1 magnetic island rotation-induced modification in gradients is found to affect the transport, flow-shear, and turbulence, providing important insights into how magnetic islands can influence broadband turbulence in the core and edge region of a tokamak [19]. Three-wave interactions between GAM and broadband electrostatic[20][21][22], as well as electromagnetic[20] turbulence in the edge plasma region, have also been observed in T-10[20][22] and Compass[21] tokamaks. An interaction between a quasi-coherent mode in density and plasma potential, in the frequency range of 50 to 200 kHz, with magnetic fluctuations has been investigated in purely Ohmic plasmas of the T-10 tokamak [23], and establishing a three-wave coupling between the quasi-coherent mode, GAM, and m/n = 2/1 MHD modes. On the other hand, the damping of GAM with the growth of m=2 MHD tearing mode, simultaneously with an increase in plasma potential, is also reported by Melnikov et al. [24]. Energy transfer through non-linear synchronization between the GAM and m/n = 6/2 MHD mode having similar frequencies in HL-2A tokamak, has also been reported. This transfer has been attributed to the location (radial) and frequency proximity of magnetic islands and GAM in the edge region of the plasma [25]. A GAM eigenmode having a similar frequency to that of magnetic fluctuations has been observed in low $q_{edge}$ ($q_{edge}$ < 2.5) discharges of SINP tokamak. It has been demonstrated that in the low $q_{edge}$ discharges, the GAM eigenmode frequency remains independent of the temperature and $q_{edge}$, and possesses non-zero poloidal mode numbers for potential fluctuations [13]. In the STOR-M tokamak suppression of a quasi-coherent GAM-like mode, characterized by m=3, 5 and n=1 magnetic components, density fluctuations with m/n=1/0, and potential fluctuations with m/n = 1/1 at a frequency ~ 35 kHz have been reported in [14].

Furthermore, it is known that magnetic islands above a particular amplitude threshold excite beta-induced Alfven eigenmodes (BAEs) [26][27][28][29][30]. Several experimental studies indicate the interactions of various coherent modes with each other as well as with the externally applied resonant magnetic perturbations (RMP). A strong non-linear interaction among EGAMs, BAEs, and tearing modes has been detected in HL-2A [31]. In the STOR-M tokamak, the excited quasi-coherent GAM-like mode has been found to be suppressed with the application of an RMP [14]. Similarly, in the MAST tokamak, suppression of coherent oscillations in density and potential fluctuations at $\approx$ 10 kHz has been observed with the application of an n = 3 resonant magnetic perturbation. It has been reported that the frequency of the coherent oscillations increases after the application of RMP, and beyond an RMP current threshold of ~ 1.4 kA, the coherent mode gets damped, coinciding with an enhancement in edge turbulence [32]. The results of multiscale interactions among zonal flows, turbulence, and magnetic islands after considering the effect of magnetic islands on turbulence, as well as the energy transfer among these from different machines have been reviewed in reference [33]. The mutual interaction between an externally induced non-rotating m/n=2/1 magnetic island (using RMP) and turbulence in KSTAR has been studied by measuring the $T_e$ profile, turbulence, and the poloidal flow using 1-D ECE and 2-D ECE imaging diagnostics. A coupling among these three parameters is observed, which has been shown to be affecting the electron heat transfer [34]. The effect of broadband magnetic fluctuations on drift wave turbulence and the associated transport has been investigated theoretically and the electromagnetic nature of drift wave turbulence has been reported [13]. However, it has been concluded that the anomalous transport is primarily dependent on electrostatic turbulence rather than magnetic fluctuations [35].

Since, the coupling of magnetic fluctuations and the edge turbulence opens channels for energy transfer [32][25] through nonlinear interactions, exploring the interaction of magnetic fluctuations and edge plasma turbulence becomes crucial for understanding various underlying dynamics leading to the interaction of tearing modes with BAE and zonal flows such as GAM. The interaction of internal MHD activity with edge plasma turbulence is an active area of research in tokamaks. In particular, an identification of the MHD excited coherent structures, that are experimentally observed, is very important for a better understanding of the underlying physical phenomena.

In this paper an experimental investigation of coupling between magnetic fluctuations and edge plasma turbulence in the high MHD plasma discharges of ADITYA-U tokamak is presented. It is shown that the MHD modes, mainly the m/n = 2/1, beyond an amplitude threshold value of $\tilde{B}_\theta/B_\theta \sim 0.3 - 0.4$ %, excite coherent oscillations in the density and plasma potential with the same frequency as that of the MHD mode. Interestingly, our detailed analyses of the excited modes reveal that the coherent mode in the edge potential fluctuation has a mode number of m / n = 2 / 1 whereas the edge density fluctuation has a mode number of m / n = 1 / 1. Beyond the threshold, the coupled power fraction between the MHD modes and density and potential fluctuations scale linearly with the magnitude of $\tilde{B}_\theta/B_\theta$ fluctuations. Furthermore, the growth rates of coherent oscillations in density and potential are also found to be dependent on the growth rate of the MHD modes; however, the mode numbers of the excited modes are found to be independent of the MHD mode frequency.

The paper is organized as follows. Section 2 presents the experimental setup in detail, including the required diagnostics and techniques used for the analysis. Section 3 describes the observed amplitude threshold of MHD activity required for the coupling between magnetic field ($\tilde{B}_\theta$) and edge plasma ($\tilde{V}_P$ and $\tilde{n}_e$) fluctuations followed by a detailed investigation of the mode structure of excited modes in density and potential. In section 4, a discussion on the plausible mechanisms that can give rise to these MHD-induced coherent oscillations in density and potential fluctuations is presented. The paper is summarized in Section 5.

## 2. Experimental setup and analysis techniques:

The experiments have been conducted in limiter discharges of ADITYA-U tokamak [36], which is an air core medium-sized tokamak with minor radius (a) = 25 cm and major radius (R) = 75 cm. The present study is carried out in repeatable Ohmically heated plasma discharges having toroidal magnetic field ($B_\phi$) = 1 T, plasma current ($I_p$) ~ 100 - 200 kA, discharge duration ~ 100 - 200 ms, chord-averaged densities of ($\bar{n}_e$) ~ $0.5 - 3 \times 10^{19} \, m^{-3}$, electron temperature ($T_e$) ~ 200 - 250 eV, edge safety factor ($q_{edge}$) ~ 3 - 5, edge density ($n_{edge}$) ~ $1$-$5 \times 10^{18} \, m^{-3}$ and edge temperature ($T_{edge}$) in the range of 5-20 eV [36]. Temporal evolution of plasma parameters for a typical discharge, such as loop voltage ($V_L$), plasma current ($I_P$), $H_\alpha$ line emission intensity ($H_\alpha$), chord average density ($n_e$), Soft X-ray intensity (SXR), poloidal magnetic fluctuations ($\tilde{B}_\theta$), and floating potential fluctuations that are a proxy for the plasma potential fluctuations $\tilde{V}_p$ measured at r=24.5 of a typical plasma discharge #36628 are shown in Figure 1. Periodic gas puffs are applied to maintain the central chord-averaged density of the plasma.

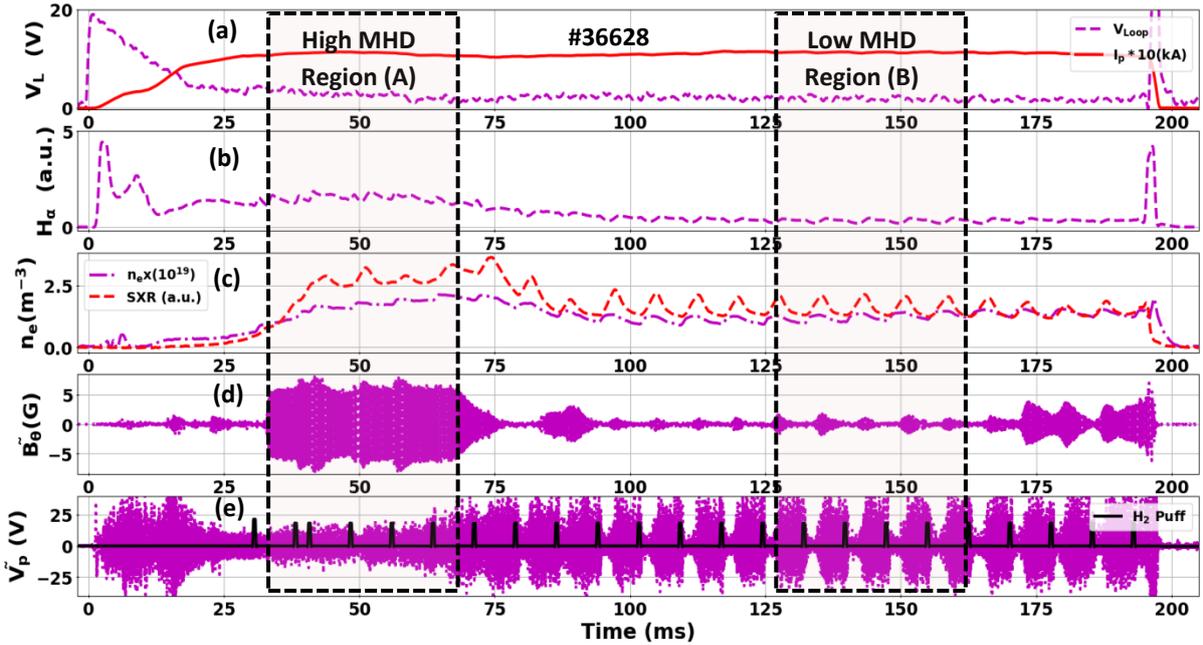

*Figure 1: Plasma parameters for discharge #36628 (a) loop voltage $V_L$ and plasma current $I_p$ ,(b) $H_\alpha$ line emission intensity (a.u.) ,(c) chord avg. density $n_e$ and Soft X-ray (SXR), (d) MHD activity $\tilde{B}_\theta$, (e) edge plasma potential fluctuation $\tilde{V}_p$, A and B represents high and low MHD region respectively.*

To control the horizontal position of the plasma column a real-time PID based feedback controller is used [37]. The poloidal magnetic fluctuations ($\tilde{B}_\theta$) are measured using two toroidally ($180^0$) separated sets of 16 Mirnov coils that are poloidally $22.5^0$ apart [38][39][40] and located at r = 28.5 cm. The data for $\tilde{B}_\theta$ is sampled at 100 kHz frequency with 16 bit digitization and analysed to obtain the nature of MHD activity ($\tilde{B}_\theta$). This analysis provides the structure of the existing magnetic islands; their rotation frequency, size, poloidal and toroidal mode numbers i.e. "m" and "n" respectively [40][41]. Figure 2 shows the location of the four sets of Langmuir probes (LPs) at different toroidal and poloidal locations installed to obtain and characterize the density and potential fluctuation in the edge region

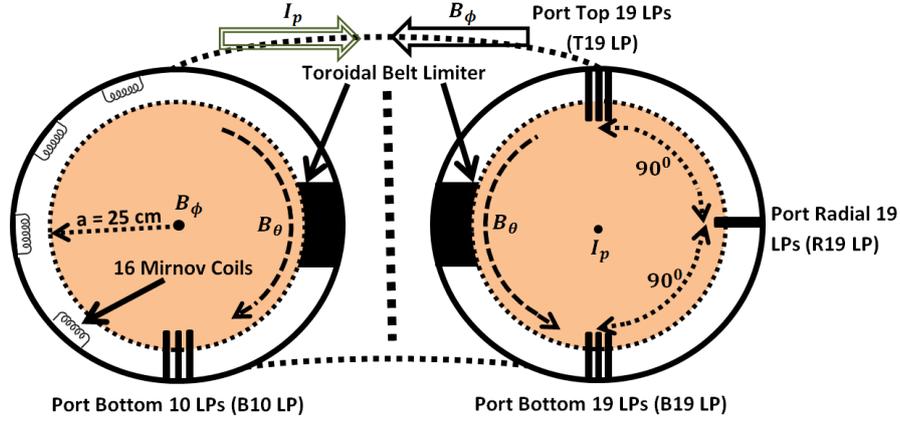

*Figure 2: Schematic of diagnostics used for the Experimental Study*

Two of the LP sets (B-19 and B-10) are installed at $160^0$ toroidally apart locations to evaluate the toroidal mode number (n) of the excited mode by analysing the data for potential fluctuations ($\tilde{V}_p$) and density fluctuations ($\tilde{n}_e$). Similarly for poloidal mode number (m) analysis, three poloidally $90^0$ separated LP sets (B19, R19, and T19) are used. All the data from LPs are recorded at a sampling frequency of 1 MHz with 14 bit digitization after a low pass filter of 100 kHz to prevent aliasing.

Potential fluctuation ($\tilde{V}_p$) and density fluctuations ($\tilde{n}_e$) are obtained after detrending the raw data obtained from the LPs [13]. Note that the temperature fluctuations are not considered while deducing the plasma potential and density fluctuations from the measurements of floating potential and ion saturation current. Magnetic fluctuations ($\tilde{B}_\theta$) are obtained after detrending the time-integrated Mirnov coil data. To quantify the coupling between the edge plasma turbulence and MHD oscillations, coherency between the $\tilde{V}_p$ and $\tilde{B}_\theta$ is calculated. Time variation of coupled power fraction ($P_{coupled}/P_{total}$; $P_{coupled}$ represents the power spectral density at the coupled frequency ($f_{MHD}$), while $P_{total}$ stands for the sum of power spectral density at all the frequencies in the frequency spectra of the normalized potential fluctuations) between edge plasma turbulence ($\tilde{V}_p$) and $\tilde{B}_\theta$ is estimated using the power spectral density (PSD) of the normalized potential ($\tilde{V}_p/V_p$) and magnetic fluctuations ($\tilde{B}_\theta/B_\theta$) for ~ 1 ms duration in intervals of ~ 0.5 ms using the Welch method [42].

Poloidal/toroidal mode numbers and coherency of excited mode ($\tilde{V}_p$, $\tilde{n}_e$ and $\tilde{B}_\theta$ coupled mode) are obtained by evaluating the cross-power spectral density (Cross-power) and coherency between poloidally/toroidally separated LPs measurements. To evaluate the Cross-power and Coherency for potential and density fluctuations, the Welch method-based "csd" and "coherence" functions from the signal module in Scipy library of Python is employed using 2048 time points for FFT with 80% overlap and the "Blackman" window giving a frequency resolution ($\Delta f$) of 0.5 $kHz$, and error (mean squared error) = Variance+Bias$^2$ = 5 % [42][43]. Mode numbers (m & n) are calculated using the measured cross-phase ($\Delta \phi$) and the distance between the probes. The cross-phase ($\Delta \phi$) is obtained from cross-power between the time-series data of two toroidally or poloidally separated probes.

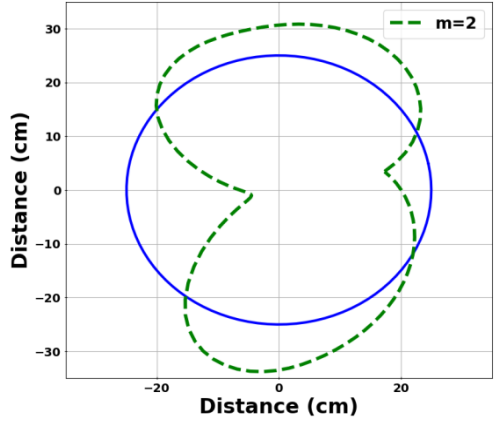

*Figure 3: Poloidal mode structure m = 2 obtained by SVD analysis for plasma discharge 36628 (60-65 ms)*

## 3. Characteristics of the MHD-induced Coherent modes in $\widetilde{V}_p$ and $\widetilde{n}_e$:

To understand the driving mechanism and characteristics of the MHD excited coherent mode in ADITYA-U tokamak, hundreds of plasma discharges are analysed. Firstly, it has been observed that there exists a threshold value of the MHD ($\widetilde{B}_\theta/B_\theta$) amplitude, above which the coherent modes in $\widetilde{V}_p$ and $\widetilde{n}_e$ are excited. Beyond the threshold, the dependencies of growth rates and power coupled to the coherent modes on the growth rates and amplitude of the MHD are obtained. Finally, the poloidal and toroidal structures of the excited mode in $\widetilde{V}_p$ and $\widetilde{n}_e$ are identified and investigated along with the frequency-wavenumber relation of the excited mode.

### (i) Coupling between Magnetic ($\widetilde{B}_\theta/B_\theta$) and edge plasma ($\widetilde{V}_p$ and $\widetilde{n}_e$) fluctuations:

In plasma discharges of ADITYA-U tokamak, the power spectral density of MHD oscillations display a typical frequency $(f_{MHD}) \approx 7 - 12 \, kHz$ corresponding to mainly m/n=2/1 magnetic island rotation (Figure 3) [39][40]. This predominant frequency in MHD activity is a measure of the rotation speed of the magnetic island while the amplitude of MHD activity ($|\widetilde{B}_\theta|/B_\theta$) is directly proportional to its width (w) given by:

$$\frac{w}{r_s} = 2\left[\left(\frac{2}{m}\right)\left(\frac{r_c}{r_s}\right)^m \left(\frac{\widetilde{B}_\theta}{B_\theta}\right)\right]^{\frac{1}{2}} \quad (1)$$

$r_s$: mode resonant radius, $r_c$: Mirnov coil location, $\widetilde{B}_\theta$ is poloidal magnetic field fluctuation, $B_\theta$ is poloidal magnetic field and m is the poloidal mode number[38][44].

As shown in figure 1, in the current flat-top region of discharge #36628, high-amplitude MHD activity is observed during 33.5 – 70 ms, whereas the MHD amplitude remains relatively low during the rest of the discharge. A zoomed time-window (31 – 35 ms) depicting the temporal evolution of normalized poloidal magnetic field fluctuation $\widetilde{B}_\theta/B_\theta$ (%), its envelope, and normalized plasma potential fluctuations $\widetilde{V}_p/V_p$ (a.u.) is shown in Figure 4(a). It can be clearly seen from the figure that during the period of low MHD, i.e., before 33.5 ms, the $\widetilde{V}_p/V_p$ shows a random behavior, whereas, when the MHD mode grows and saturates at relatively higher amplitude ($|\widetilde{B}_\theta|/B_\theta \geq 0.3$ %) after 33.5 ms, the amplitude of potential fluctuations also increases and they start oscillating at the same frequency of the MHD mode.

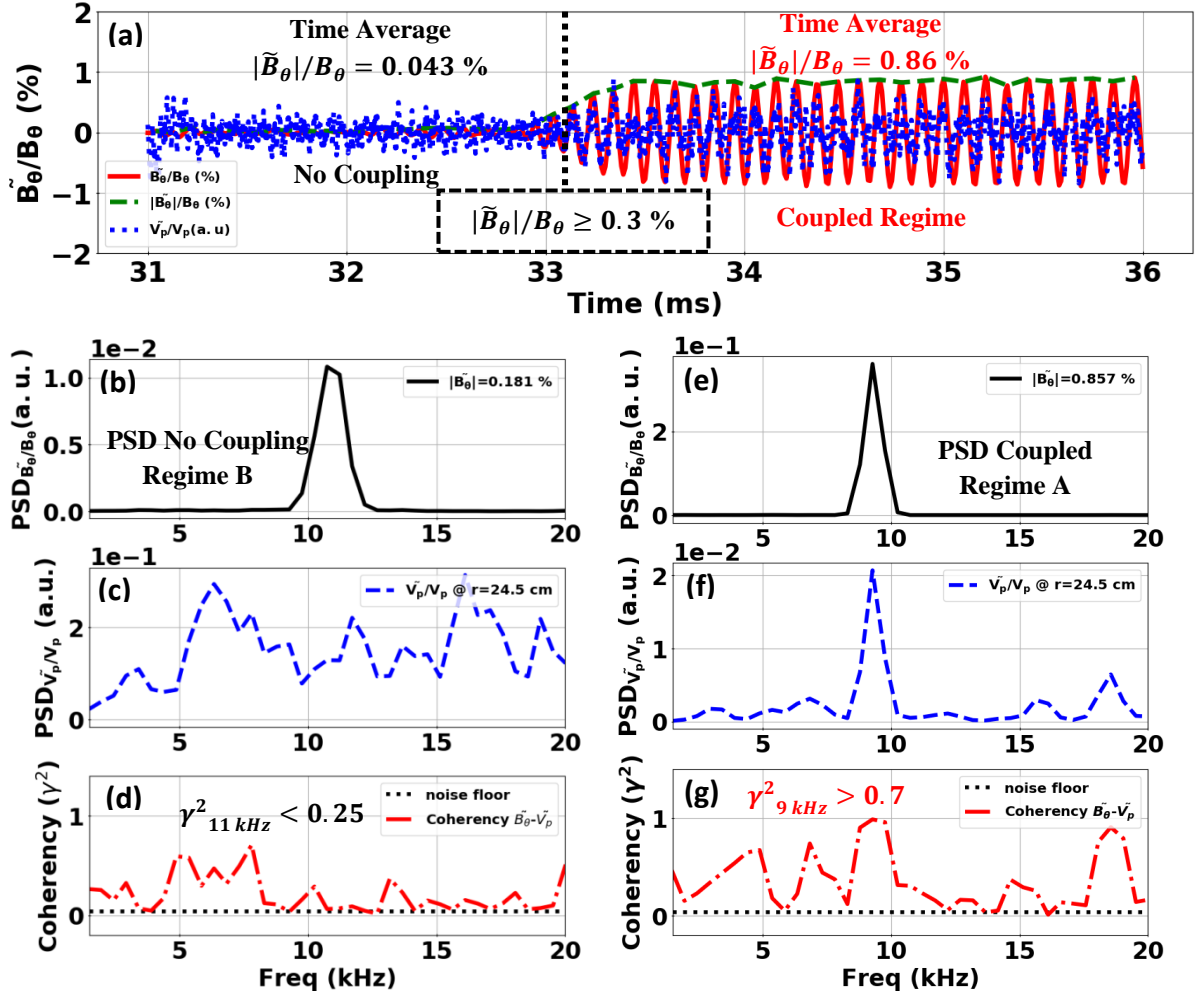

*Figure 4:* (a) Modulation in $\tilde{V}_p$ due to high amplitude of $\tilde{B}_\theta$ for plasma discharge #36628 (b) power spectral density for $\tilde{B}_\theta/B_\theta$ (c) power spectral density for $\tilde{V}_p/V_p$, (d) Coherency between $\tilde{B}_\theta$ and $\tilde{V}_p$ for Low MHD activity region $|\tilde{B}_\theta|/B_\theta = 0.181$ % in red dash-dot line, black dotted line represents the noise floor in coherency estimation; (e) power spectral density for $\tilde{B}_\theta$ (f) power spectral density for $\tilde{V}_p$, (g) Coherency between $\tilde{B}_\theta$ and $\tilde{V}_p$ for high MHD activity region $|\tilde{B}_\theta|/B_\theta = 0.86$ % in red dash-dot line, black dotted line represents the noise floor in coherency estimation; for plasma discharge #36628

For a quantitative analysis, two time-windows, ~ 33.5 to 70 ms (marked as region A in figure 1) and 130 to 160 ms (marked as region B in figure 1) are chosen where high and low amplitude MHD activity are present, respectively. Fast Fourier Transform (FFT) analysis of the time series data of $\tilde{B}_\theta/B_\theta$ and $\tilde{V}_p/V_p$ has been carried out during these time intervals. Figure 4 (b, c, d) and Figure 4 (e, f, g) represent the power spectral density (PSD) of ($\tilde{B}_\theta/B_\theta$), PSD of ($\tilde{V}_p/V_p$) and coherency ($\tilde{B}_\theta - \tilde{V}_p$) for the two regions "B" and "A" respectively. In low MHD regime (average $|\tilde{B}_\theta|/B_\theta$ ~ 0.18 %), a dominant frequency peak at ~10 kHz can be clearly seen in the frequency spectra of $|\tilde{B}_\theta|/B_\theta$ as shown in Figure 4b while the frequency spectra of $\tilde{V}_p/V_p$ depicts broadband nature with no coherent peaks (Figure 4c). The coherency between $\tilde{B}_\theta$ and $\tilde{V}_p$ ($\tilde{B}_\theta - \tilde{V}_p$) is also observed to be < ~ 0.25 at MHD mode frequency ~ 11 kHz (Figure 4d), indicating no coupling between the magnetic fluctuations and edge plasma turbulence. Note here that no coupling between $\tilde{B}_\theta$ and $\tilde{V}_p$ was observed outside the region "A" as the MHD activity remained low almost throughout the entire plasma current flat-top phase of discharge #36628 as shown in figure 1. In region A, where the MHD amplitude is relatively high (average $|\tilde{B}_\theta|/B_\theta \sim 0.8\,\%$), a dominant frequency peak at ~ 9 kHz is observed in the PSD of both the $\tilde{B}_\theta/B_\theta$ and $\tilde{V}_p/V_p$ (Figure 4e and 4f). The coherency ($\tilde{B}_\theta - \tilde{V}_p$) value ($\gamma^2$) of > 0.7 at ~ 9 kHz indicates a coupling between the $\tilde{B}_\theta$ and $\tilde{V}_p$ (Figure 4g). A peak around 18 kHz is also detected in the coherency estimation between magnetic and potential fluctuations which may be indicative of the existence of a second harmonic of the mode. Furthermore, relatively higher coherency (~ 0.5) is also observed in the frequency region 4 kHz < f < 8 kHz, which is found to be independent of the amplitude of the magnetic fluctuations $|\tilde{B}_\theta|/B_\theta$ (Figure 4d and 4g); however this phenomena is not observed consistently in many of the plasma discharges used for the study.

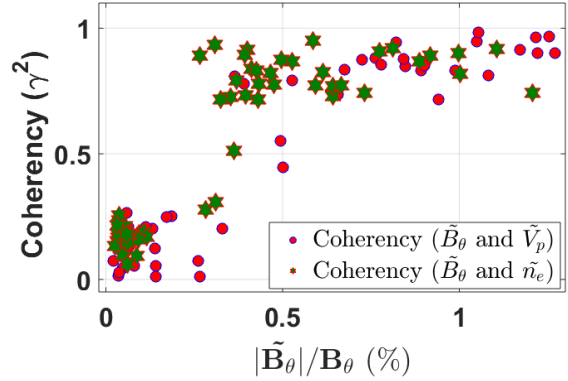

*Figure 5: Statistical analysis showing variation of coherency corresponding to MHD mode frequency between magnetic ($\tilde{B}_\theta$) and edge plasma fluctuation ($\tilde{V}_p$ and $\tilde{n}_e$) with $|\tilde{B}_\theta|/B_\theta$(%)*

The coherency between $\tilde{B}_\theta$ and $\tilde{V}_p$ ($\tilde{B}_\theta - \tilde{V}_p$), $\tilde{B}_\theta$ and $\tilde{n}_e$ ($\tilde{B}_\theta - \tilde{n}_e$) as a function of the amplitude of $\tilde{B}_\theta/B_\theta$ corresponding to the MHD mode frequency, obtained by analyzing ~ 100 ADITYA-U discharges, is plotted in Figure 5. Figure 5 shows the existence of a threshold amplitude of $\tilde{B}_\theta$ ($|\tilde{B}_\theta|/B_\theta \geq 0.3 - 0.4\%$), above which the potential and density fluctuations in the edge region are found to be modulated by the poloidal magnetic field fluctuations in ADITYA-U tokamak. As shown in Figure 5, below the threshold, the coherency values corresponding to the MHD mode frequency of ($\tilde{B}_\theta - \tilde{V}_p$), and ($\tilde{B}_\theta - \tilde{n}_e$) is < 0.25 whereas it is > 0.7 beyond the threshold. The data presented in Figure 5 is from discharges with constant toroidal magnetic field, $B_\phi \sim 1$ T. Note that the threshold value of $|\tilde{B}_\theta|/B_\theta$ for the coupling between $\tilde{B}_\theta$ and $\tilde{V}_p/\tilde{n}_e$ as a function of toroidal magnetic field ($B_\phi$) is also explored and found to be independent on $B_\phi$. The ion-saturation current fluctuations that are a proxy for the edge density fluctuations also exhibit exactly similar behavior as that of the potential fluctuations in terms of the observation of coherent modes at higher MHD amplitude (Figure 6), existence of threshold (Figure 5) and the coherency between $\tilde{B}_\theta$ and $\tilde{n}_e$ (Figure 5 and 6).

This analysis indicates the existence of a critical threshold required for coupling of magnetic fluctuations and edge plasma turbulence in ADITYA-U tokamak. Moreover, in case of MHD activity below the threshold ($|\tilde{B}_\theta|/B_\theta < 0.3\%$), the mode analysis of the FFT-filtered time-series data of the edge $\tilde{n}_e$ and $\tilde{V}_p$ from all the three probes (top, radial and bottom) in a frequency range where the dominant MHD mode lies, did not show any coherent spatial structures. The mode numbers obtained are completely random for the data analyzed from several discharges. The mode numbers obtained are completely random and cphase 1+ cphase2 = cphase 3 (as discussed in sec 3 ii) did not remain valid for the data analyzed from several discharges. Note that the coherent modes in $\tilde{V}_p$ and $\tilde{n}_e$ are always observed with the high MHD activity beyond the threshold value mentioned above.

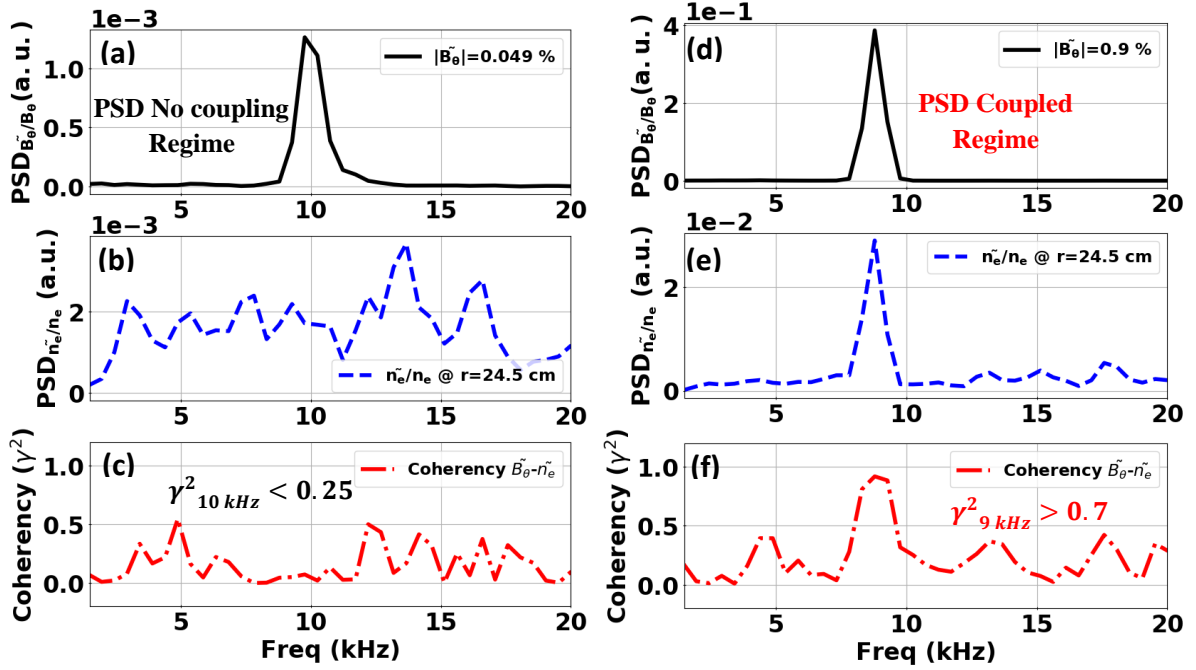

*Figure 6: (a) power spectral density for $\tilde{B}_\theta/B_\theta$ (b) power spectral density for $\tilde{n}_e/n_e$ ,(c)Coherency between $\tilde{B}_\theta$ and $\tilde{n}_e$ for Low MHD activity region $|\tilde{B}_\theta|/B_\theta = 0.049$ % ; (d) power spectral density for $\tilde{B}_\theta$ (e) power spectral density for $\tilde{n}_e/n_e$ ,(f) Coherency between $\tilde{B}_\theta$ and $\tilde{n}_e$ for high MHD activity region $|\tilde{B}_\theta|/B_\theta = 0.9$ % for plasma discharge #36671*

Further analysis revealed that beyond the threshold the coupled power fraction defined by $P_{coupled}/P_{total}$; ($P_{coupled}$ represents the power spectral density at the coupled frequency ($f_{MHD}$), while $P_{total}$ stands for the sum of power spectral density at all the frequencies in the frequency spectra of the normalized potential fluctuations) between the $\tilde{B}_\theta$ and $\tilde{V}_p$ or $\tilde{n}_e$ increases almost linearly with an increase in amplitude of $\tilde{B}_\theta/B_\theta$ as shown in Figure 7a. The rise-rate of the coherent modes in $\tilde{V}_p$ and $\tilde{n}_e$ is also found to be dependent on the growth rate of the MHD mode. The variation of rise rate of coupled power ($\gamma(PSD_{coupled})$; $PSD_{coupled} = P_{coupled}$ ) with growth rate of MHD mode ($\gamma(|\tilde{B}_\theta|/B_\theta)$) is plotted in Figure 7b and is seen to be increasing linearly. The data and analyses presented above establish the conclusion that when the MHD mode-amplitude increases beyond a certain threshold value, a coupling between magnetic fluctuations and edge plasma fluctuations occurs. To understand further the characteristics of these MHD driven coherent modes, the spatial structures of these modes in poloidal and toroidal directions are obtained and analyzed.

**(ii) Poloidal structure of the excited mode:**

Four sets of poloidally and toroidally separated LPs located inside the last closed flux surface (LCFS) at a radius, r = 24.5 cm, have been used to measure the potential and density fluctuations at different poloidal and toroidal locations. These measurements are used to obtain the poloidal and toroidal mode structures of the excited mode. Three poloidally separated, bottom, radial and top probe sets located at one toroidal location are represented by B19, R19, and T19 respectively. Another probe set is inserted from the bottom port (B10) at another toroidal location ~ 160 degree away from the location of poloidally separated probes. After detrending the time series data, potential and density fluctuations from the poloidally separated probes are analyzed for cross-power, cross-phase, and coherency to calculate the poloidal mode number (m) with total time segments of 5 and 10 ms for the analysis. Mean values of edge density and plasma potential are compared in Figure 8a and 8f respectively for plasma discharge #36671 and #36628, where the highlighted region represents the MHD-induced mode excitation window. Figures 8b and 8g show the poloidal phase difference in $\tilde{n}_e$ and $\tilde{V}_p$ for the excited mode frequency at ~ 9 kHz. Cross power among poloidally separated measurements is represented in Figures 8c and 8h for $\tilde{n}_e$ and $\tilde{V}_p$ respectively.

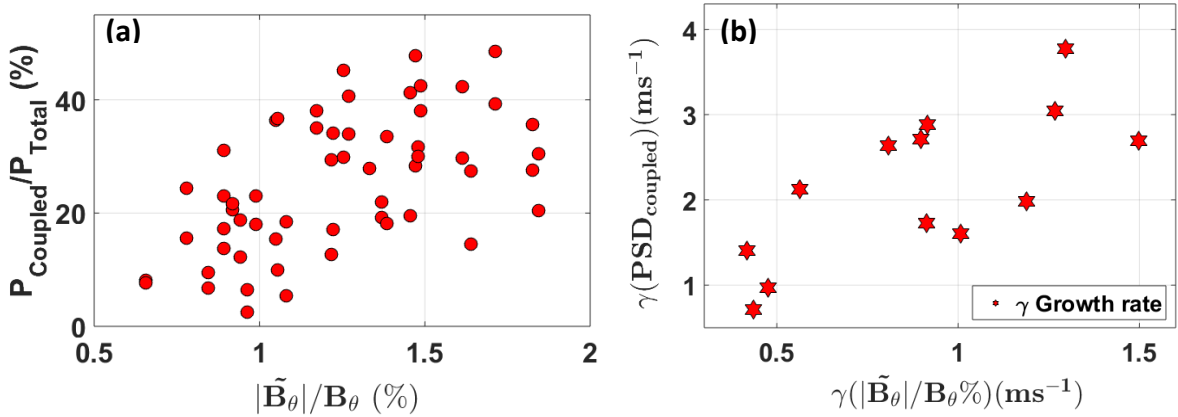

*Figure 7:* (a) Statistical analysis representing the variation of coupled power fraction with $|\tilde{B}_\theta|/B_\theta$ (b) Statistical variation of rise rate of coupled power with growth rate of $|\tilde{B}_\theta|/B_\theta$

The phase difference for $\tilde{n}_e$ between the measurements from probes placed $90^0$ apart i.e. B-19 and R-19 (Cphase-1) and R-19 and T-19 (Cphase-2) are estimated to be ~ 1.5 rad with a shot-to-shot statistical variation of $\pm 0.2$ rad. The phase difference between the probes which are $180^0$ apart i.e. B-19 and T-19 (Cphase-3) is estimated to be ~ $3\pm 0.2$ rad. The cross phases of poloidally separated probes for $\tilde{I}_s$ (#36671) are presented in Figure 8d. The data clearly indicate the existence of a highly coherent mode with coherency > 0.8 (Figure 8e), which follows the relation;

$$\text{Cphase-1} + \text{Cphase-2} \approx \text{Cphase-3}$$

The poloidal mode number (m) for $\tilde{n}_e$ from these calculations comes out to be ~ $0.9 \pm 0.1$ from the calculations of phase difference using cross-power as illustrated in detail in section 2.

Similar analysis for $\tilde{V}_p$ has been carried out and is presented in Figures 8h, 8i, and 8j for plasma discharge #36628. For $\tilde{V}_p$, the measured Cphase-1 comes out to be ~ $3.09 \pm 0.2$, the Cphase-2 comes out to be ~ $-2.8 \pm 0.2$ and Cphase-3 is ~ $0.38 \pm 0.2$ rad. Following the relation: Cphase-1 + Cphase-2 ≈ Cphase-3, it indicates an existence of a coherent mode with Coherency > 0.7 (Figure 8 (j)) with a poloidal mode number ~ $1.8 \pm 0.2$.

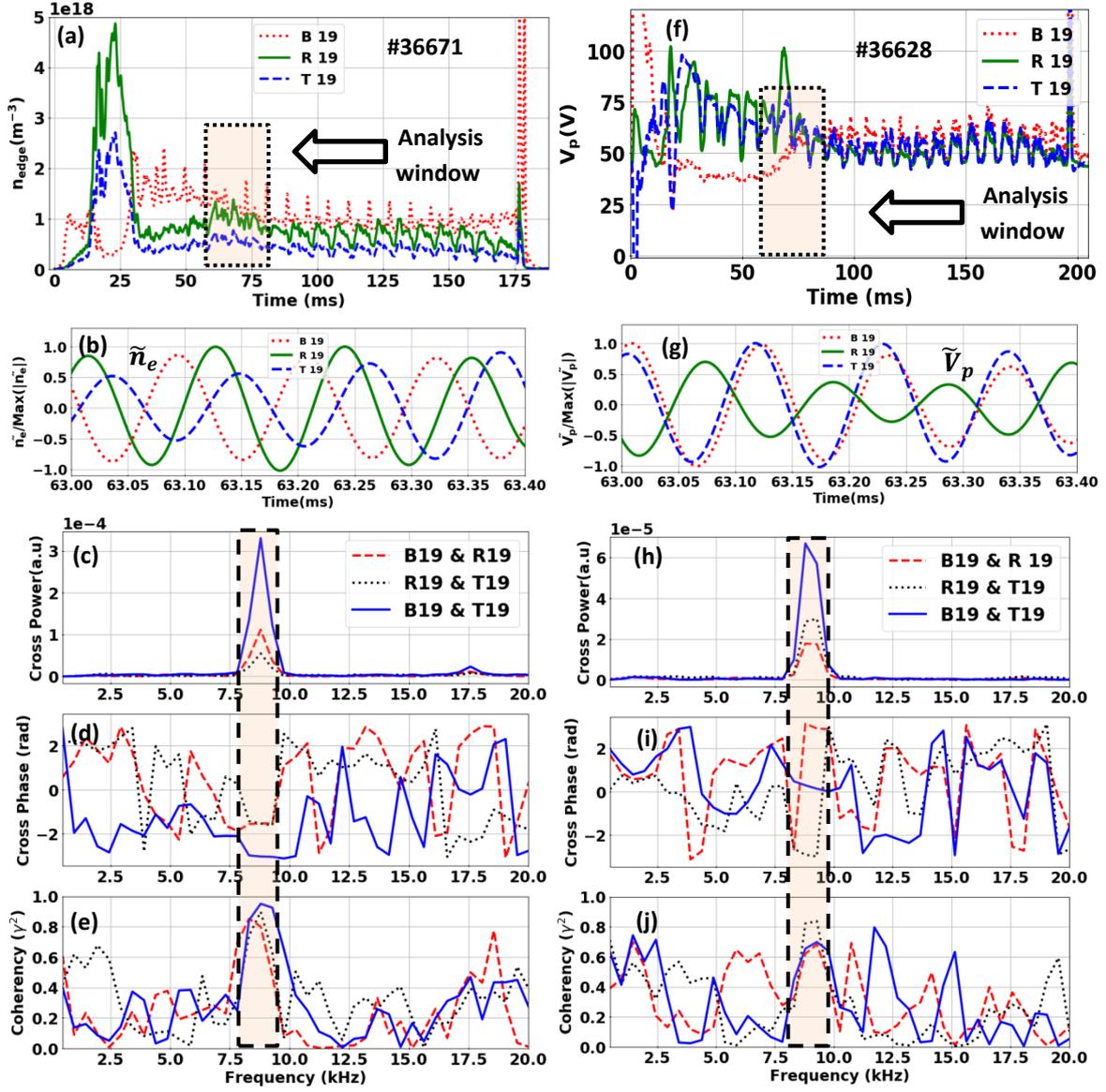

*Figure 8:* (a) edge density (r= 24.5 cm) for plasma discharge #36671, (b) phase difference among poloidaly separated measurement of $\tilde{n}_e$ (c) poloidal Cross-power,(d) cross-phase, and (e) coherency for $\tilde{n}_e$ (60-65 ms) indicating poloidal mode number (m) for $\tilde{n}_e \approx 1$; (f) edge plasma potential (r= 24.5 cm) for plasma discharge #36628, (g) phase difference among poloidaly separated measurement of $\tilde{V}_p$, (h) poloidal Cross-power,(i) cross-phase, and (j) coherency for $\tilde{V}_p$ (60-65 ms) indicating poloidal mode number (m) for $\tilde{V}_p \approx 2$.

### (iii) Toroidal structure of the excited mode:

The toroidal mode numbers of coherent modes of $\tilde{V}_p$ and $\tilde{n}_e$ are obtained from measurements made at two toroidally separated LPs (B-19 and B-10) located at the same poloidal and radial (r = 24.5 cm) positions using similar analysis as explained in previous subsection. Figure 9a shows the toroidal cross power peaking at ~ 9 kHz. The toroidal phase difference for $\tilde{n}_e$ is observed to be 2.6 $\pm$ 0.2 rad, and for $\tilde{V}_p$ it is 2.1 $\pm$ 0.2 rad as represented in Figure 9b which is also evident from Figure 9d and 9e where the time series data of $\tilde{V}_p$ and $\tilde{n}_e$ in the frequency window of 9 $\pm$ 2 kHz is plotted from both the probes. Analysis of toroidal measurement of $\tilde{n}_e$ and $\tilde{V}_p$ reveals a coherent mode with coherency > 0.8 (Figure 9c) corresponding to the toroidal mode numbers (n) ~ 0.9 ± 0.1 and ~ 0.8 ± 0.1 respectively.

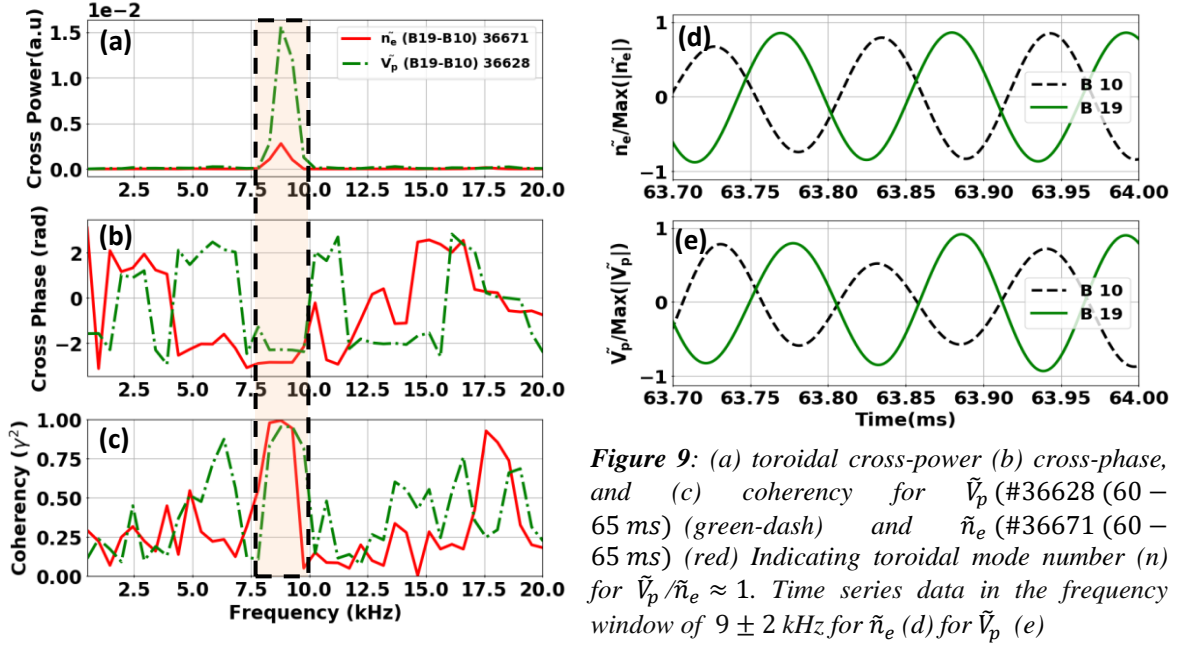

*Figure 9*: *(a) toroidal cross-power (b) cross-phase, and (c) coherency for $\tilde{V}_p$ (#36628 (60 − 65 ms) (green-dash) and $\tilde{n}_e$ (#36671 (60 − 65 ms) (red) Indicating toroidal mode number (n) for $\tilde{V}_p/\tilde{n}_e \approx 1$. Time series data in the frequency window of $9 \pm 2$ kHz for $\tilde{n}_e$ (d) for $\tilde{V}_p$ (e)*

**(iv) Frequency-Wavenumber relation of the excited modes:**

To investigate further the nature of MHD induced coherent modes in $\tilde{V}_p$ and $\tilde{n}_e$, the statistical variation of $k_\theta$ (poloidal wave number) is plotted as a function of frequency of coherent modes in $\tilde{n}_e$ and $\tilde{V}_p$ in figure 10a and 10b respectively. In the coupled regime, a variation of MHD mode frequency in the range of 8 – 10 kHz is observed from different plasma discharges analysed for this study. Correspondingly, the frequency of coherent modes in $\tilde{n}_e$ and $\tilde{V}_p$ also varied in the range of 8 – 10 kHz. To obtain the $\omega$-k relationship, the wavenumber ($k_\theta$) estimated from the mode number ($m = k_\theta r$), is plotted against the coherent frequency in the available frequency range. It can be clearly seen from Figure 10, that the wavenumbers of the coherent modes in both $\tilde{n}_e$ and $\tilde{V}_p$ remained independent of the mode frequency in the range of ~ 8 – 10 kHz.

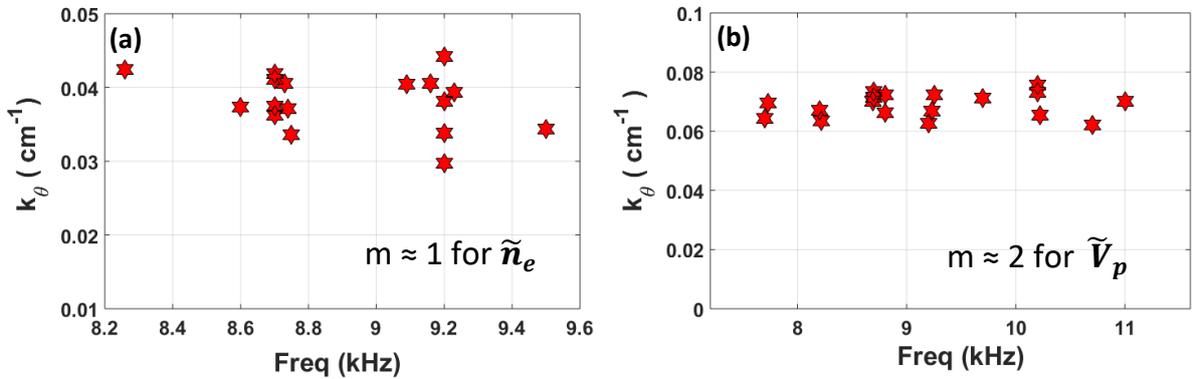

*Figure 10*: *wavenumber (cm$^{-1}$) vs frequency (kHz) plot (a) for $\tilde{n}_e$ and (b) for $\tilde{V}_p$*

The poloidal mode number (m) has a shot–to–shot variation of 0.9 ± 0.1 and 1.8 ± 0.2 for $\tilde{n}_e$ and $\tilde{V}_p$ respectively. To assess the accuracy of the values for m=1 and m=2, the standard error ($\sigma/\sqrt{n}$) is calculated from multiple plasma discharges as well as from different ensembles of the same plasma discharge. For m=1, the standard error calculated from the standard deviation is ~ 0.022, corresponding to a percentage error of approximately 2%. For m=2, the standard error comes out to be ~ 2.35%.

## 4. Discussion:

The observations and analyses presented above establish that in typical discharges of ADITYA-U tokamak, whenever the amplitude of the m/n =2/1 resistive MHD mode, increases beyond a threshold value of ($|\tilde{B}_\theta|/B_\theta \geq 0.3 - 0.4$ %), coherent modes with the same frequency as that of $\tilde{B}_\theta$, appear in the $\tilde{n}_e$ and $\tilde{V}_p$, measured in the edge plasma region.

Below the threshold value of $\tilde{B}_\theta$, the edge plasma remains predominantly turbulent with no coherent modes. This indicates that there exist a critical threshold value of $\tilde{B}_\theta$, above which the resistive MHD modes affect and modify the edge plasma turbulence in ADITYA-U tokamak. The experimentally observed MHD induced electrostatic coherent modes can be physically understood as a synchronization (or frequency entrainment) phenomenon between a driver wave (MHD mode) and a particular electrostatic mode in the broadband spectrum of the background turbulence that matches the frequency of the driver. Synchronization is an inherently nonlinear phenomenon that is dependent on the amplitude of the driver. In driving and amplifying the electrostatic mode the driver has to overcome any intrinsic dissipation (damping) that the mode has and this determines the threshold amplitude of the driver.

It has also been observed that, beyond the threshold, the coupled power fraction in $\tilde{n}_e$ and $\tilde{V}_p$ increases almost linearly with the amplitude of $\tilde{B}_\theta$ and the growth rate of the MHD mode also dictates the rise rate of the coupled power in $\tilde{n}_e$ and $\tilde{V}_p$. Although, the correlations between electrostatic and magnetic fluctuations in the edge region have been earlier detected and reported in tokamaks[11][15][16][45], here we show that measurements of the spatial structures of the coupled modes in edge $\tilde{n}_e$ and $\tilde{V}_p$ provide important insights into the nature of these coherent modes. One curious feature that emerges from the mode number measurements is that the coherent mode in $\tilde{V}_p$ has an m/n = 2/1 structure, which is same as that of the $\tilde{B}_\theta$, but, the coherent mode in $\tilde{n}_e$ has an m/n = 1/1 structure. Furthermore, the wavenumber ($k_\theta$) is found to be independent of the frequency of the resistive MHD mode.

Several possible mechanisms may be considered for explaining the above-mentioned observations. They include the coupling of drift waves with the Alfven wave (drift-Alfven waves)[46][47][48], excitation of Beta-induced Alfven Eigen modes (BAE) [26][27][28][29] resulting from strong MHD activity, the coupling of drift waves with resistive tearing modes (drift-tearing mode)[49][40][39], zonal flows driven by the magnetic islands etc. In the following, arguments in favour and against each of these mechanisms in explaining the observations are discussed.

The observed coherent mode in the edge $\tilde{n}_e$ and $\tilde{V}_p$ could be due to the excitation of a Drift-Alfven wave (DAW), resulting from the coupling of drift waves and Alfven waves[46][47]. Such a coupling is most likely when the parallel drift wave velocity, $\omega/k_\parallel$, becomes comparable to the Alfven velocity, $v_A$, $\omega/k_\parallel \sim v_A$. In our case, with $B_\phi \sim$ 1T, $n_e \sim 2 \times 10^{19} m^{-3}$, the Alfven velocity $v_A \sim 5 \times 10^6 \, m/s$ is much higher than $\omega/k_\parallel \sim 2 \times 10^4 \, m/s$, with $\omega \sim 2\pi \times 10^4 Hz$ and $k_\parallel^{-1} \sim qR \sim 1.5 \, m$. Furthermore, the m/n = 2/1 mode structure of the high-amplitude (beyond threshold) MHD modes and their relatively slower growth rates indicate that these are resistive MHD modes. This suggests that the observed coherent modes in the edge $\tilde{n}_e$ and $\tilde{V}_p$ are less likely to be due to the excitation of a DAW.

An instability with a frequency lower than the Alfven eigen mode and with low mode numbers, identified as a beta-induced Alfven eigen (BAE) mode, has been observed in a wide variety of tokamak plasmas[26][27][28][29]. This mode is represented as a pair of counter-propagating, tearing-parity waves that are excited in the presence of a magnetic island and has the same helicity as the magnetic island. The BAEs interaction with drift waves can lead to coherent modes in edge $\tilde{n}_e$ and $\tilde{V}_p$. In the FTU tokamak excitation of BAE with a frequency ~ 30-70 kHz was reported in discharges with poloidal magnetic fluctuations above a threshold amplitude of ~ 0.2 % [27][28]. The low frequency ($f < 50\ kHz$) quasi coherent modes observed in Ohmic as well as NBI assisted L and H-mode plasma discharges of compass tokamak are identified as BAE using linear MHD code KINX with m, n < 5 [50]. Quasi-coherent oscillations in the same frequency range observed in the Ohmic discharges with magnetic islands in T-10 [51] and TEXTOR[52] tokamaks were also identified as BAEs. Furthermore, the frequency of the BAEs was found to be proportional to the magnetic island width ($|\tilde{B}_\theta|/B_\theta$) and dependent on the toroidal magnetic field. The lowest order BAE frequency is given by $\omega_{BAE} = \frac{1}{R_0}\sqrt{\frac{2T_i}{m_i}\left(\frac{7}{4} + \frac{T_e}{T_i}\right)}$ [28]. Taking the parameters of ADITYA-U tokamak's $H_2$ plasma discharges with $T_e/T_i \approx 3$ and $T_i \approx 50\ eV$ at the island location ($r_s \approx 18\ cm$), the lowest order frequency of BAEs comes out to be ~ 45 kHz. This frequency of the lowest order BAE is also much higher than the frequency of MHD excited coherent mode in edge $\tilde{n}_e$ and $\tilde{V}_p$ observed in our experiments. More importantly, in our experiments, the frequency of the MHD excited coherent modes in edge $\tilde{n}_e$ and $\tilde{V}_p$ is observed to be inversely proportional to the amplitude of $\tilde{B}_\theta$ as shown in Figure 11a and is also independent of the toroidal magnetic field (Figure 11b). Therefore, the observed coherent modes in the edge $\tilde{n}_e$ and $\tilde{V}_p$ in our experiments could not be a result of a drift-BAE interaction. Such a mechanism would also not explain observed mode numbers of m/n = 1/1 in $\tilde{n}_e$.

The so called drift-tearing modes [38][39][40][49], are excited when drift modes couple to tearing modes in the regime of high diamagnetic-drift frequency. The electron response to the magnetic perturbation of the island drives the electron drift wave at a frequency $0 < \omega(k_\perp) < \omega_e^*$ with the $k_\perp$ of the separatrix layer resonating with the electron drift wave having frequencies in the range $0 < \omega(k_\perp) < \omega_e^*$. Drift-tearing modes remain an attractive possibility for interpreting the observed coherent modes in the edge $\tilde{n}_e$ and $\tilde{V}_p$ in our experiments. They can certainly provide support for the observation of the m/n = 2/1 in potential fluctuations[53], but they cannot explain the observation of m/n = 1/1 mode structure in the density fluctuations.

Another possible mechanism for driving the observed MHD-excited coherent mode in edge $\tilde{n}_e$ and $\tilde{V}_p$ may be the MHD induced zonal flows. Several experimental [6][54] observations and theoretical[7] work have demonstrated the existence of self-generated zonal flows in tokamaks. It has been reported that the fishbone (a macroscopic MHD mode) induced zonal flows are likely to be responsible for the formation of an internal transport barrier in DIII-D discharges [55]. Furthermore, the modification in mean zonal radial electric field ($E_r$) has been explored in several theoretical and experimental investigations [56][57][58]. Recent simulations also report the existence of m/n = 2/1 potential structures induced by m/n = 2/1 magnetic islands [53].

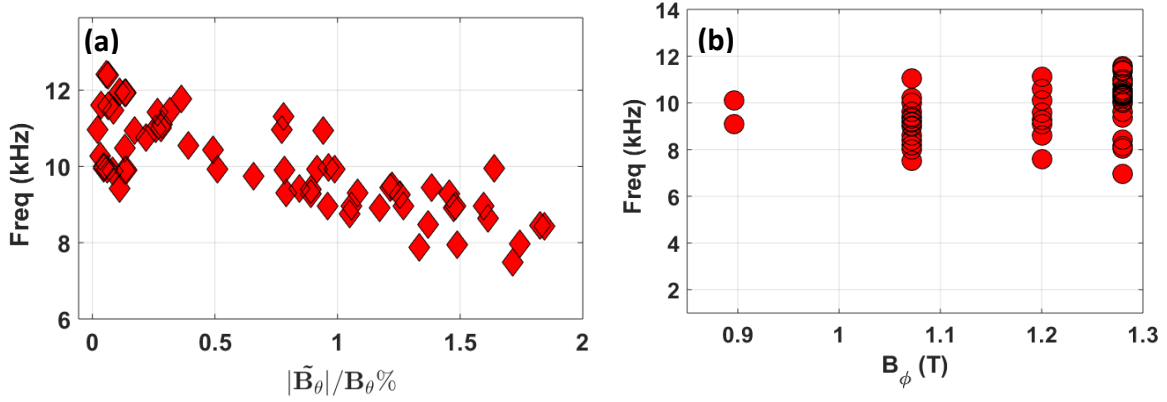

*Figure 11*: (a) *frequency (kHz) vs $|\tilde{B}_\theta|/B_\theta$* (b) *frequency (kHz) vs toroidal magnetic field ($B_\phi$) (T)*

The existence of the odd poloidal modes (m=1) in density fluctuations, due to even poloidal mode in potential fluctuations (m=2), can be explained by the 1/R dependence of toroidal magnetic field owing to the toroidicty of the tokamak. The radial electric field fluctuations ($\tilde{E}_r$), originating due to the $\tilde{V}_p$, leads to a $\tilde{E}_r \times B_\phi$ drift resulting in a poloidally asymmetric velocity fluctuation due to the radial (major) variation of toroidal magnetic field ($B_\phi$). The poloidally asymmetric velocity fluctuations generate the m / n = 1 / 1 mode in the density fluctuation. Since, there is no such asymmetry in toroidal direction, the toroidal mode structure for $\tilde{V}_p$ and $\tilde{n}_e$ remain identical to the driving magnetic oscillations (n=1 for $\tilde{B}_\theta$, $\tilde{V}_p$ and $\tilde{n}_e$). This is a well-established paradigm for explaining the observation of m=1 mode in density fluctuations due to m=0 mode in potential fluctuations occurring in the geodesic acoustic mode (GAM) excitation. Furthermore, the toroidicity providing a coupling of even harmonics of the potential with odd harmonics of the pressure, and vice versa is also theoretically well founded as reported in reference [59].

Note that the frequency ~ 10 kHz of the observed MHD-induced coherent modes in $\tilde{n}_e$ and $\tilde{V}_p$ in our experiments lies near the high-frequency branch of zonal flows, although it does not depend on the local plasma temperature as in the case of global eigenmode GAMs. This hypothesis is further supported by the possible excitation of high-frequency branch of zonal flows i.e. GAMs through application of external magnetic perturbation in tokamaks[60]. However, the conventional GAM is characterised by m/n=0/0 mode in potential and m/n=1/0 in the density fluctuations, whereas our observations show m/n = 2/1 in potential and m/n =1/1 in the density fluctuations indicating the excitation of a coherent mode bearing similar characteristics to the high frequency zonal flows which may be termed as GAM-like.

## 5. Summary and Conclusion:

In typical discharges of ADITYA-U tokamak, coherent modes in the edge density and plasma potential fluctuations, having the same frequency that of the MHD mode, are observed whenever the amplitude of MHD activity increases beyond a threshold value, $|\tilde{B}_\theta|/B_\theta \sim 0.3 - 0.4$ %. The mode-number and growth-rate analysis reveals that the MHD activity is associated with the existence of m/n = 2/1 resistive tearing modes. Interestingly, the Langmuir probe measurements from different radial, poloidal, and toroidal locations in the edge plasma region demonstrate m/n = 1/1 and m/n=2/1 spatial structures of the MHD generated coherent modes in density and potential fluctuations, $\tilde{n}_e$ and $\tilde{V}_p$, respectively. The observations of linearly increasing coupled power fraction to the $\tilde{n}_e$ and $\tilde{V}_p$ coherent modes with the magnitude of $\tilde{B}_\theta/B_\theta$ fluctuations along with strong dependence of the rise rates of the coupled power fraction on the growth rate of $\tilde{B}_\theta/B_\theta$ fluctuations further establish the significant coupling between the magnetic, $\tilde{B}_\theta$ and electrostatic fluctuations $\tilde{n}_e$ and $\tilde{V}_p$. These observations are found to be highly repeatable and do not vary with the toroidal magnetic field in the range of 1 – 1.4 T. Furthermore, the frequency-wavenumber relation exhibit that the wavenumber ($k_\theta$) remains independent in the frequency range of 8 – 10 kHz. Although, similar observations of the correlations between electrostatic and magnetic fluctuations in the edge region have been reported earlier[11][15][16][45], the mode-number measurements of the coherent modes in $\tilde{n}_e$ and $\tilde{V}_p$ fluctuations, presented in this paper, are novel and throw new light towards better understanding of the possible causes. Out of the many possible physical causes for explaining the above-mentioned observations, such as the excitation of drift-Alfven waves, drift-BAE interaction, drift-tearing modes etc., the global high frequency branch of zonal flows occurring through coupling of even harmonics of potential to the odd harmonics of pressure due to 1/R dependence of toroidal magnetic field owing to the toroidicty of the tokamak seems to best support our experimental observations.

**Acknowledgment and Author Contributions:**

The work reported in this paper comprises the Ph.D. dissertation research of the first author, who is registered in HBNI, Mumbai. The first author is responsible for conceptualizing, designing, and conducting the experiments, constructing the probes, analysing the data, preparing the original draft, and editing later versions. The first author is the primary contributor to the work. For the fabrication of Langmuir probes, the authors are thankful to the IPR- workshop and drafting section. AS is grateful to the Indian Science Academy for the INSA Honorary Scientist position.

**Data Availability**

The data that support the findings of this study are available from the corresponding author upon reasonable request.